\documentclass[preprint,aps,floatfix,superscriptaddress,pre,showpacs]{revtex4}
\usepackage{amsmath, amsthm, amssymb, graphicx}
\input epsf.sty
\begin{document}

\def\g{\gamma}
\def\r{\rho}
\def\w{\omega}
\def\wo{\w_0}
\def\wp{\w_+}
\def\wm{\w_-}
\def\t{\tau}
\def\av#1{\langle#1\rangle}
\def\pf{P_{\rm F}}
\def\pr{P_{\rm R}}
\def\F#1{{\cal F}\left[#1\right]}

\title{Aging processes in systems with anomalous slow dynamics}

\author{Nasrin Afzal and Michel Pleimling}
\affiliation{Department of Physics, Virginia Tech, Blacksburg, Virginia 24061-0435, USA}

\date{\today}

\begin{abstract}
Recently, different numerical studies of coarsening in disordered systems have shown the existence of a crossover
from an initial, transient, power-law domain growth to a slower, presumably logarithmic, growth.
However, due to the very slow dynamics and the long lasting transient regime, 
one is usually not able to fully enter the asymptotic regime when investigating the relaxation of these systems
toward equilibrium. We here study two simple
driven systems, the one-dimensional $ABC$ model and a related domain model with simplified
dynamics, that are known to exhibit anomalous slow relaxation where the asymptotic
logarithmic growth regime is readily accessible. Studying two-times correlation
and response functions, we focus on aging processes and dynamical scaling during logarithmic growth.
Using the time-dependent growth length as the scaling variable, a simple aging picture emerges that
is expected to also prevail in the asymptotic regime of disordered ferromagnets and spin glasses.
\end{abstract}

\maketitle

\section{Introduction}
Recent years have seen remarkable progress in our understanding of physical aging in non-disordered
systems with slow, i.e. glassy-like, dynamics (see \cite{book10} for a recent comprehensive overview).
In many systems, ranging from ferromagnets undergoing phase-ordering \cite{Bra94} to reaction-diffusion
systems \cite{Hen07}, a single dynamical length $L(t)$, that grows as a power-law of time $t$,
governs the dynamics out of equilibrium. In the aging or dynamical scaling regime these systems
are best characterized by two-times quantities, like dynamical correlation and response functions,
that transform in a specific way under a dynamical scale transformation \cite{Cug03}. The resulting dynamical
scaling functions and the associated non-equilibrium exponents are
often found to be universal and to depend only on some global features
of the system under investigation.

However, growth laws can be much more complicated, as discussed recently in disordered ferromagnets quenched below their
critical temperature.  Thus,
convincing evidence for a dynamic crossover
between a transient regime, characterized by a power-law growth with an effective dynamical exponent that depends on the
disorder, and the asymptotic regime, where the growth is logarithmic in time, has been found in recent studies of
the dynamics of elastic lines
in a random potential \cite{Kol05,Noh09,Igu09,Mon09} as well as in numerical simulations of disordered 
Ising models \cite{Rao93,Aro08,Par10,Cor11,Cor11a,Cor12,Par12}.
These indications are compatible with the classical (droplet) theory of activated dynamics that,
under the assumption of energy barriers growing as a power of $L$, predicts a slow
logarithmic increase \cite{Hus85} of this length: 
\begin{equation}
L \sim  (\ln t )^{1/\psi}, 
\end{equation}
with the barrier exponent $\psi > 0$.
Whereas in some of the studies on disordered Ising models aging phenomena in the crossover regime were investigated
\cite{Aro08,Par10,Cor11,Cor11a,Cor12,Par12},
none of these recent numerical studies was able to enter so deeply into the asymptotic regime
that no corrections to the logarithmic growth law were detectable any more.
Therefore a systematic study
of aging processes in this regime with pure logarithmic growth has not yet been done.

In this paper we study two one-dimensional models that exhibit anomalous slow dynamics and that are known to display
coarsening where the length of the domains increases logarithmically with time \cite{Eva02}. Even though these models are in no way
related to disordered ferromagnets and spin glasses, their studies should allow us to gain a better understanding
of the generic properties of an aging system with a logarithmic growth law.

The models discussed in the following are the so-called $ABC$ model \cite{Eva98a}, a driven 
diffusive system composed of three different types of
particles that swap places asymmetrically, and a related domain model \cite{Eva98b} whose simplified dynamics is supposed to
capture the dynamics of the $ABC$ model at the later stages of the coarsening process.
The $ABC$ model has recently yielded a flurry of interesting 
studies \cite{Cli03,Bod08,Ayy09,Led10a,Led10b,Bar11a,Ber11,Bar11b,Ger11,Coh11,Bod11,Ger12,Coh12,Coh12b} that helped
establishing it as a paradigm for systems far from equilibrium. Not only is the $ABC$ model characterized
by its anomalous slow dynamics, making it a representative for a larger class of systems with a similar coarsening process
\cite{Lah97,Arn98,Lah00,Kaf00,Lip09}, it also exhibits a variety of interesting non-equilibrium phase transitions
whose properties change dramatically when breaking certain conservation laws. The domain model has been proposed 
as a simplified version of the $ABC$ model where only movements of particles between domains of the some species are considered.
This simplified dynamics accelerates the coarsening process and allows to enter the purely logarithmic growth
regime faster \cite{Eva98b}. In the following we use the $ABC$ model in order to investigate the onset of dynamical scaling,
whereas the domain model is used to characterize aging scaling deep inside the logarithmic growth regime.

Our paper is organized as follows. In the next Section we discuss in more detail the two models that we study.
Section III is devoted to the aging processes taking place in the $ABC$ model. We thereby
focus on the two-times autocorrelation function where the two times are not always in the asymptotic, logarithmic
scaling regime. In
Section IV we characterize aging scaling in the domain model through the study of both correlation and
response functions. We discuss our results in Section V.

\section{Models and quantities}

In the $ABC$ model particles of three different species live on a one-dimensional ring \cite{Eva98a}. 
Every lattice site is occupied by exactly one particle, which can swap places with its
left and right neighbors. In the symmetric case, where all exchanges happen with the same rate,
every particle undergoes a random walk, and nothing interesting takes place. However, this
changes dramatically as soon as one introduces a bias which makes the particles diffuse
asymmetrically around the ring. This is achieved by randomly selecting a pair of neighboring sites
and updating them using the following rates:
\begin{eqnarray} \label{eq:ABC_rates}
AB \overset{q}{\underset{1}{\rightleftarrows}} BA \nonumber \\
BC \overset{q}{\underset{1}{\rightleftarrows}} CB \nonumber \\
CA \overset{q}{\underset{1}{\rightleftarrows}} AC 
\end{eqnarray}
with $q < 1$.
As a result of these rules, phase separation takes place in such a way that the ordered domains
arrange themselves in
repetitions of the sequence $ABC$, where $A$ indicates a domain
of $A$ particles, followed by a domain of $B$ particles, which itself is followed by a $C$ domain. 
Once this arrangement has been achieved, the domains coarsen whereby the
typical domain size increases logarithmically with time. 

Obviously these exchanges keep constant the total number of particles of each species. We consider
in our study only lattice sizes divisible by three and initially populate one third of the lattice sites
by particles of each species. In that case detailed balance is fulfilled and the system evolves
toward an equilibrium steady state \cite{Eva98a}.

In the domain model one focuses on the later stages of the coarsening process where well-defined, compact
domains have already formed. One then defines a simplified dynamics where only events are taken into account 
that change the sizes of two neighboring domains of the same species. For example, consider the case where 
two such $A$ domains are selected, called $A_l$ and $A_r$, that are separated by one $B$ and one $C$ domain,
yielding the sequence $\cdots A_lBCA_r \cdots$. Calling $a_l$ respectively $a_r$ the domain size of the
domain $A_l$ respectively $A_r$, these domain sizes
are then modified in one of the following two ways \cite{Eva98b}:
\begin{eqnarray}
a_l \longrightarrow a_l -1 ~~, ~~ a_r = a_r +1 ~~~\mbox{with rate}~q^b \nonumber \\
a_l \longrightarrow a_l + 1 ~~, ~~ a_r = a_r -1 ~~~\mbox{with rate}~q^c \nonumber
\end{eqnarray}
where $b$ respectively $c$ are the number of sites of the $B$ respectively $C$ domain separating
our two $A$ domains. These rates follow from the observation that in order to go from one domain to the other an $A$ particle
has to cross one of the two intermediate domains in the `wrong' direction. The domain model therefore exclusively
considers processes where particles successfully travel between domains of the same type, irrespective on how 
many jumps are needed for that transit.

Two-times quantities are well suited to study relaxation processes far from equilibrium \cite{book10}. 
We here briefly recall the expected behavior of such quantities, without entering into the details on how these
quantities are computed for our driven diffusive systems. This will be done in the following Sections when we
discuss our numerical results.

The two-times quantities usually at the center of aging studies are the autocorrelation function $C(t,s)$ and the autoresponse
function $R(t,s)$. The autocorrelation function measures the extend to which configurations taken at two different times
$s$ and $t > s$ are correlated. Here $s$ is the waiting time, whereas $t$ is called the observation time. The autoresponse
function, on the other hand, allows us to investigate how the system reacts during the relaxation process
to a instantaneous perturbation (as for many other studies, we will focus below on the time integrated response
to a longer lasting perturbation which is much easier to measure). In the aging
regime, where the observation and waiting times are large compared to any microscopic time scale, 
the single growth length $L$ dominates the properties of the system, so that the different quantities
should depend on time only through this length $L$. Thus one expects the following
(very general) scaling forms, using standard notation \cite{book10}:
\begin{eqnarray}
C(t,s) & = & \left(L(s)\right)^{-b}  f_C\left(\frac{L(t)}{L(s)}\right) \label{eq:C_scaling}\\
R(t,s) & = & \left(L(s)\right)^{-1 -a} f_R\left(\frac{L(t)}{L(s)}\right)\label{eq:Rscaling} 
\end{eqnarray}
with the scaling functions $f_C(y)$ and $f_R(y)$ and the non-equilibrium exponents $a$ and $b$. In systems
undergoing coarsening one usually has $b = 0$ and $a \neq 0$, but this can be different in other situations,
as for example during non-equilibrium relaxation at a critical point \cite{Cal05}.
In cases with an algebraic growth law $L(t) \sim t^{1/z}$, as observed in critical systems or coarsening systems
without disorder, one usually uses $t/s$ as scaling variable. However, for more complicated cases with subleading
contributions to the growth and/or crossover between an initial algebraic growth and the true asymptotic
behavior, this approach is too simplistic and $L(t)/L(s)$ has to be used as variable in order to achieve
the expected scaling \cite{Par10,Par12}.

\section{Aging in the $ABC$ model}
In our simulations of the original $ABC$ model we focus on the early time regime
where coarsening slowly sets in. We thereby always prepare the system
in a disordered initial state with every species occupying one third of the lattice sites chosen at random.
The data presented below have been obtained for rings with $N = 9000$ sites. This is large enough so that no finite 
size effects show up for the times accessed in our simulations, as we checked by making additional runs for other
system sizes. We define one time step as $N$ proposed updates. For every proposed update we select a pair of neighboring 
sites at random and then exchange them with the rates given in (\ref{eq:ABC_rates}).

\subsection{Domain growth}
\begin{figure} [h]
\includegraphics[width=0.80\columnwidth]{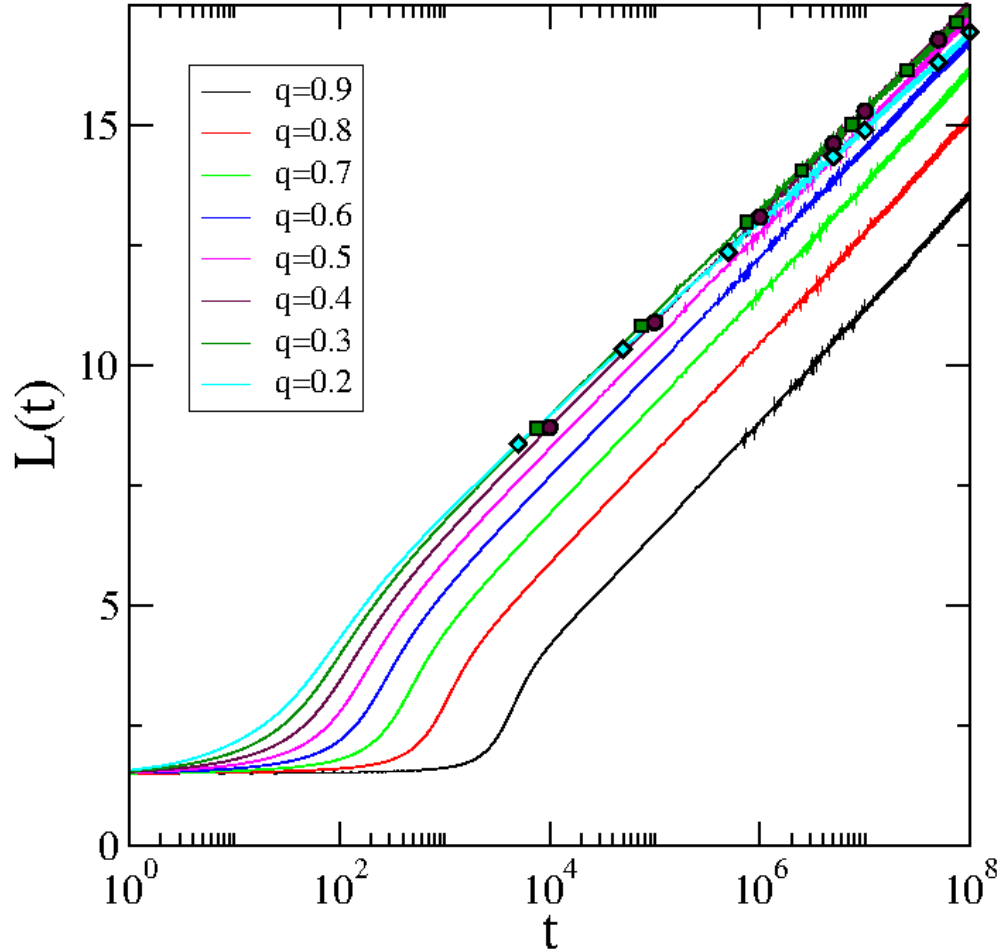}
\caption{\label{fig1} (Color online)
Time-dependent average domain size for the $ABC$ model with various values of the rate $q$. After some initial regime,
that lasts longer the larger the value of $q$, logarithmic growth sets in. The slopes in the log-linear plot
increase with $q$. The data result from averaging over 600 independent runs. For small $t$ values, the curves are ordered
in such a way that the largest $q$ value corresponds to the lowest curve, whereas the smallest $q$ value yields the
highest curve. For larger $t$ the curves start to cross, due to difference in slopes. 
In order to make this crossing better visible, some selected 
data points are shown as symbols (circles: $q=0.4$, squares: $q=0.3$, diamonds: $q=0.2$).
}
\end{figure}

We start by having a look at the average domain size. Fig. \ref{fig1} shows $L(t)$ for a large range of $q$ values.
We note that in all cases an initial regime is observed during which domains are formed and arranged in the
correct sequence, so that a $C$ domain follows a $B$ domain that follows an $A$ domain. This initial regime
lasts longer for larger values of $q$ as it gets increasingly difficult to form these initial domains the closer
$q$ gets to 1. 

%%%%%%%%%%%%%%%%%%%%%%%%%%%%%%%%%%%%%%%%%%%FIG 1.%%%%%%%%%%%%%%%%%%%%%%%%%%%%%%%%%%%%%%%%%%%%%%%%%%%%%%
%%%%%%%%%%%%%%%%%%%%%%%%%%%%%%%%%%%%%%%%%%%FIG 1.%%%%%%%%%%%%%%%%%%%%%%%%%%%%%%%%%%%%%%%%%%%%%%%%%%%%%%

Once these initial domains are formed, they then coarsen, and the system size increases logarithmically with
time: $L ( t) \sim \ln t$. Obviously, this is a very slow process and even after $10^8$ time steps the average domain size
does not reach twenty lattice spacings. This coarsening proceeds faster for
larger values of $q$. Indeed, the slopes in the log-linear plot decrease when decreasing $q$. Thus, in the
interval between $t=10^6$ and $t=10^8$ we obtain that the slope continuously decreases from 1.05 for $q=0.9$
to 0.86 for $q=0.2$. Whereas at short times the domain size is the largest for the smallest $q$ value, we expect
the order to be reversed for very long times, due to the difference in slopes. In fact, indications 
of this are already seen in Fig.\ \ref{fig1}, see the two curves for $q=0.2$ and $q=0.3$ that start to be
below some of the curves obtained for larger $q$ values.

A closer inspection of the curves in Fig.\ \ref{fig1} for the smallest $q$ values 0.2 and 0.3
reveals that their slopes change slightly with time. 
Even after $t = 10^8$ time steps we are for these $q$ values not yet completely inside the asymptotic regime where corrections to the logarithmic growth
law should be completely absent. 

\subsection{Autocorrelation}

As mentioned in the introduction, valuable insights into relaxation far from equilibrium can be gained through
the study of two-times quantities. In this subsection we discuss the autocorrelation $C(t,s)$. For our three species
system we characterize lattice site $i$ by a time-dependent Potts variable $p_i(t)$(alternatively we could use a
species dependent occupation number \cite{Afz11,Ahm12}) that can take on the three different values 0, 1, or 2, depending
on whether at time $t$ the site is occupied by an $A$, $B$, or $C$ particle. The autocorrelation function $C(t,s)$ is then defined 
as 
\begin{equation} \label{eq:C}
C(t,s) = \left< \frac{1}{N} \sum\limits_{i=1}^N \delta_{p_i(t),p_i(s)} \right> - \frac{1}{3}
\end{equation}
where $\delta_{\alpha,\beta}$ is the Kronecker delta. In that equation $\left< \cdots \right>$ indicates
an average over both initial conditions and noise as realized through different random number sequences.
We subtract from this average the value 1/3 that one has for two completely uncorrelated configurations,
thus making sure that $C(t,s)$ approaches zero when t gets very large.

In our simulations we averaged over a large number of realizations, ranging from 600 for the longest waiting
times to 20000 for the shortest waiting times. In all cases we let the system evolve for $t = 40~s$ time
steps where $s$ is the waiting time.

%%%%%%%%%%%%%%%%%%%%%%%%%%%%%%%%%%%%%%%%%%%FIG 2.%%%%%%%%%%%%%%%%%%%%%%%%%%%%%%%%%%%%%%%%%%%%%%%%%%%%%%
\begin{figure} [h]
\includegraphics[width=0.80\columnwidth]{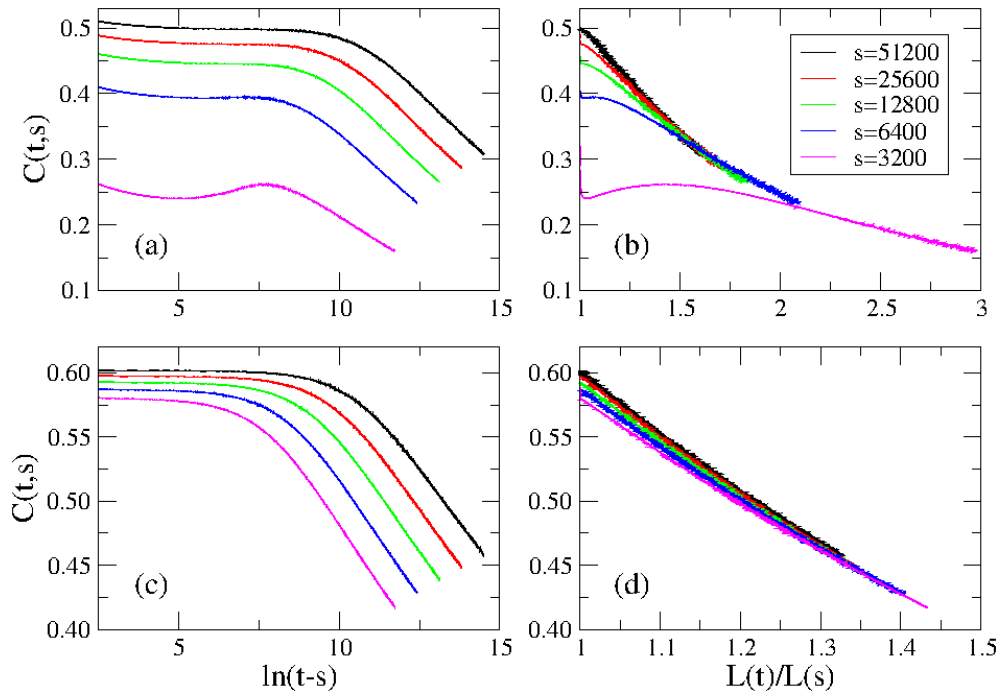}
\caption{\label{fig2} (Color online)
Autocorrelation function for the $ABC$ model with (a,b) $q=0.9$ and (c,d) $q=0.3$. For every 
waiting time $s$ we compute the autocorrelation function for up to $t = 40~s$ time steps.
Plotting the autocorrelation
against the scaling variable $L(t)/L(s)$, see (b) and (d), yields indications for the onset of dynamical
scaling for the longest waiting times. The data result from averaging over at least 600 independent runs.
The values of the autocorrelation increase with increasing waiting times.
}
\end{figure}
%%%%%%%%%%%%%%%%%%%%%%%%%%%%%%%%%%%%%%%%%%%FIG 2.%%%%%%%%%%%%%%%%%%%%%%%%%%%%%%%%%%%%%%%%%%%%%%%%%%%%%%

The data shown in Fig. \ref{fig2} for $q = 0.9$ and $q = 0.3$ are representative for all studied values of $q$.
Comparing data for different waiting times reveal the expected physical aging where the two-times
quantity is not simply a function of the time difference, see Fig. \ref{fig2}a and \ref{fig2}b. For $q=0.9$ the 
behavior for the shortest waiting time shown in Fig. \ref{fig2}a clearly differs from that observed for the larger
waiting times. In fact, inspection of Fig.\ \ref{fig1} reveals that $s=3200$ lies in the time regime where the initial
domains are forming and where coarsening starts to set in. As a result correlations dramatically change in the system, which 
is revealed by the non-monotonous behavior of the autocorrelation function.

In Fig. \ref{fig2}b and \ref{fig2}d we test dynamical scaling by plotting the data as a function of
$L(t)/L(s)$. Clear deviations are observed for the smaller waiting times, but these deviations get less and less 
important the larger $s$ gets, yielding for $q=0.3$ already a good data collapse for the largest waiting times.
All this indicates that for very large $s$ we start to be in the aging scaling regime. In agreement with
Fig.\ \ref{fig1} the scaling regime is accessed more rapidly for the smaller $q$ values. We also note that
even for $t/s = 40$, the ratio of the corresponding lengths $L(t)/L(s)$ remains rather small. Obviously,
the regime $L(t)/L(s) \gg 1$ remains out of reach in systems displaying logarithmic growth.

\section{Aging in the domain model}

It follows from the discussion in the previous Section that it is extremely difficult to fully enter the
asymptotic growth regime for the $ABC$ model. We therefore focus in the following on the domain model
with simplified dynamics that captures the essential properties of the $ABC$ model deep inside the
coarsening regime while speeding up the dynamics \cite{Eva98b}.

For the domain model we consider systems with $N=27000$ sites, thereby checking carefully that no finite-size
effects affect our data for the times accessed in our simulations. As the dynamics assumes the
existence of domains that coarsen, we prepare our system in an initial state where we have 3000 sequences
of $ABC$ domains, with every domain extending over three lattice sites. We then start the system with 
the chosen value of $q$. During the simulations smaller domains tend to disappear as larger domains
keep growing. If, say, an $A$ domain vanishes in the original $ABC$ model, this yields a sequence $ABCBCA$, 
which rapidly evolves into a sequence $ABCA$ as for two neighboring sites $CB$ is replaced by $BC$
with rate 1. The resulting $B$ respectively $C$ domains have then sizes that are identical to the sums of the sizes
of the two $B$ respectively $C$ domains at the moment of the dismissal of the $A$ domain. In the domain model this merging is done
immediately whenever a domain vanishes \cite{Eva98b}. For simplicity we increase in our simulations time $t$ by one unit
when the number of proposed updates is equal to the number of domains that are in the system at time
$t$. 

\subsection{Domain growth}

\begin{figure} [h]
\includegraphics[width=0.80\columnwidth]{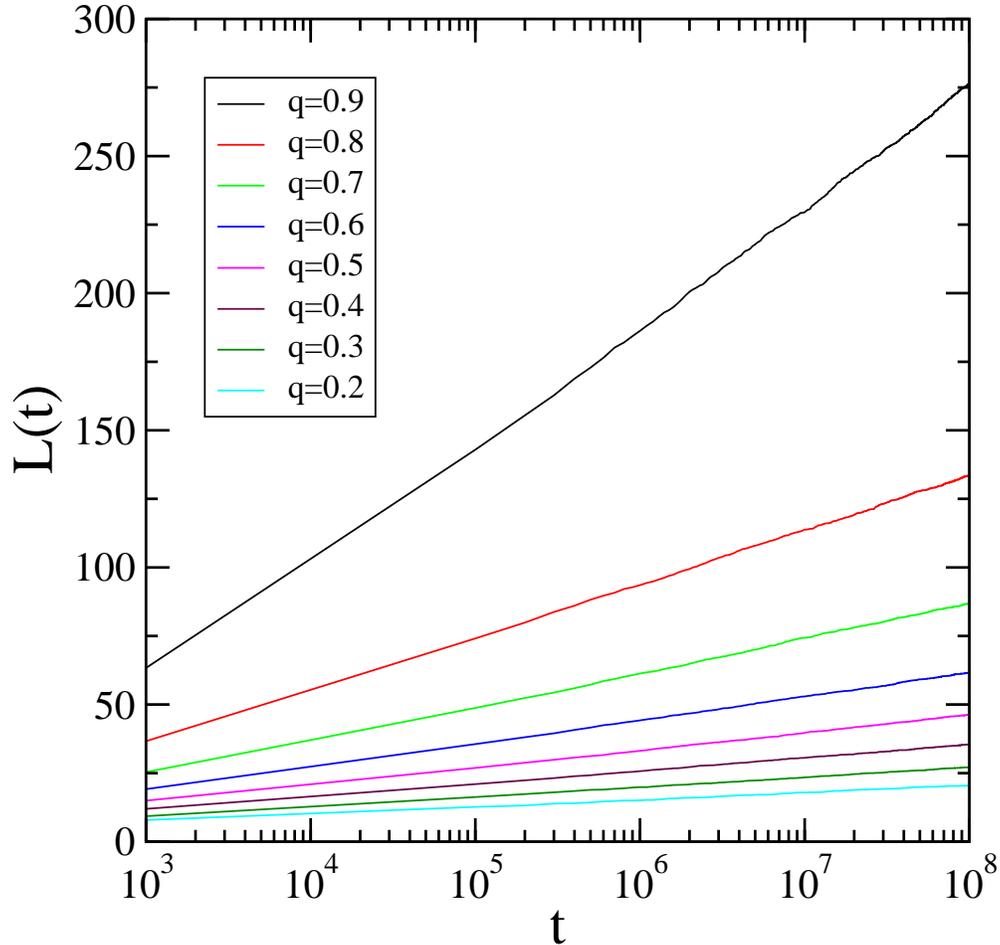}
\caption{\label{fig3} (Color online)
Time-dependent average domain size for the domain model for various values of the rate $q$. Logarithmic
growth is observed where the slopes in the log-linear plot increase with $q$. 
The data result from averaging over at least 100 independent runs.
For a fixed time $t$ the domain size is larger the larger the value of the rate $q$ is.
}
\end{figure}

In Fig. \ref{fig3} we verify that we are indeed deep inside the logarithmic growth regime for all
studied values of $q$. As already observed in \cite{Eva98b}, the logarithmic growth sets in very
rapidly when using the simplified dynamics. We note that the growth proceeds faster for larger
values of $q$. This is of course in agreement with our observation in Fig.\ \ref{fig1} that 
for the system with the full dynamics the
prefactor in the equation (which corresponds to the slope in the log-linear plot)
\begin{equation}
L(t) = \gamma \ln t
\end{equation}
is decreasing when $q$ decreases. In \cite{Eva98b} it has been proposed that the length should
grow as
\begin{equation}
L(t) = p \ln t / | \ln q|~
\end{equation}
for the domain model.
We indeed obtain consistently a value of $p \approx 2.0$ for all $q$ values. This value is slightly smaller
than the value of 2.6 found in \cite{Eva98b}. This difference should be due to the different definitions of
a time step in both studies.

%%%%%%%%%%%%%%%%%%%%%%%%%%%%%%%%%%%%%%%%%%%FIG 3.%%%%%%%%%%%%%%%%%%%%%%%%%%%%%%%%%%%%%%%%%%%%%%%%%%%%%%
%%%%%%%%%%%%%%%%%%%%%%%%%%%%%%%%%%%%%%%%%%%FIG 3.%%%%%%%%%%%%%%%%%%%%%%%%%%%%%%%%%%%%%%%%%%%%%%%%%%%%%%

\subsection{Autocorrelation}

\begin{figure} [h]
\includegraphics[width=0.80\columnwidth]{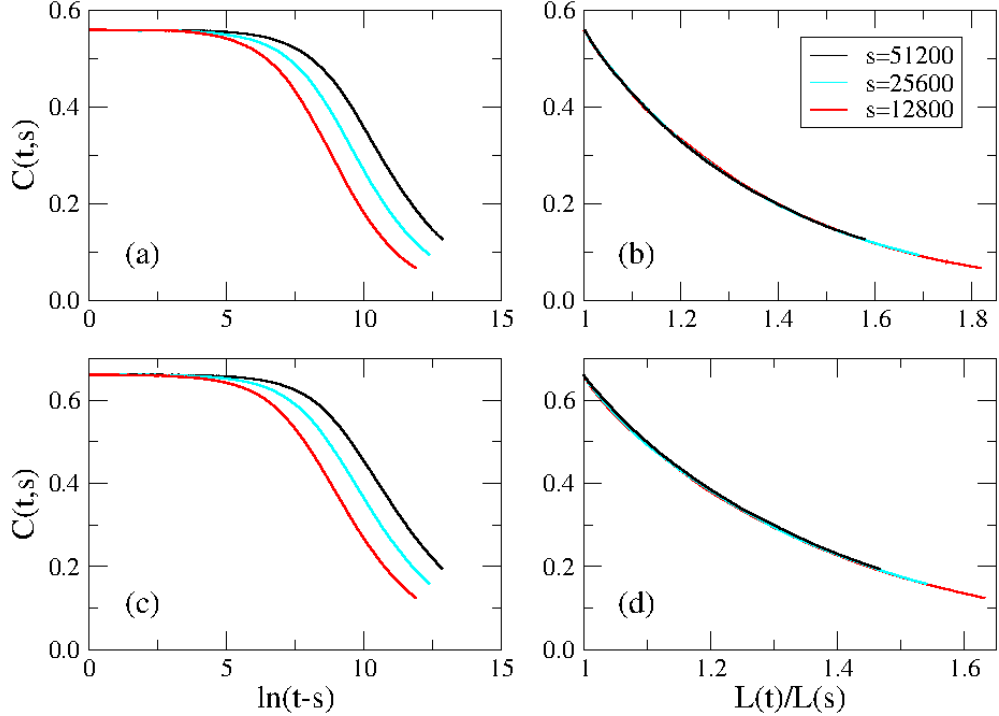}
\caption{\label{fig4} (Color online)
Autocorrelation function for the $ABC$ model with (a,b) $q=0.9$ and (c,d) $q=0.7$. 
Plotting the autocorrelation
against the scaling variable $L(t)/L(s)$, see (b) and (d), yields a perfect data collapse.
The data result from averaging over 50000 independent runs.
}
\end{figure}

For the autocorrelation we proceed as for the original $ABC$ model. Using Eq. (\ref{eq:C}) we compute 
$C(t,s)$ for various waiting times $s$ and plot the data as a function of $L(t)/L(s)$. The
result is shown in Fig. \ref{fig4} for two values of $q$. In all cases we achieve perfect data collapse
when plotting the data in this way, see Fig. \ref{fig4}b and \ref{fig4}d. This vindicates the simple
aging scaling form (\ref{eq:C_scaling}) also for systems with anomalous slow dynamics. As for
the autocorrelation only configurations at different stages of the time evolution are
compared, we expect to
encounter for that quantity the same scaling in other systems characterized by a single length scale 
that grows logarithmically with time, including disordered ferromagnets and spin glasses in their
asymptotic regime.

%%%%%%%%%%%%%%%%%%%%%%%%%%%%%%%%%%%%%%%%%%%FIG 4.%%%%%%%%%%%%%%%%%%%%%%%%%%%%%%%%%%%%%%%%%%%%%%%%%%%%%%
%%%%%%%%%%%%%%%%%%%%%%%%%%%%%%%%%%%%%%%%%%%FIG 4.%%%%%%%%%%%%%%%%%%%%%%%%%%%%%%%%%%%%%%%%%%%%%%%%%%%%%%

\subsection{Different responses}

Changes in the relaxation process due to external perturbations are best captured through the study of
two-times response functions. For spin systems, as for example ferromagnets or spin glasses, 
one of the often used protocols, both in theoretical \cite{book10} and experimental \cite{Vin06,Muk10} studies,
consists of applying a (random) magnetic field at the moment of a temperature quench. This field
is then removed after the waiting time and the relaxation of the system is monitored.

For the domain model we employ a similar scheme for the computation of the response. Preparing
the system in the same way as for the calculation of the autocorrelation, we let the system initially evolve
with a given exchange rate $q = q_i$. At time $t=s$ we change the exchange rate to its final value $q = q_f$
that is kept constant until the end of the run. Due to the initial value of $q$, the average domain size at
the waiting time $s$ differs from the typical domain size encountered in a system that evolves at
the fixed value $q=q_f$. Consequently we choose as our observable the difference in system sizes between the
perturbed system, where we switch from $q_i$ to $q_f$, and the unperturbed system, where $q=q_f$ for the whole run:
\begin{equation} \label{eq:M}
M(t,s) = \left| L_p(t,s) - L(t) \right|~.
\end{equation}
Here $L_p(t,s)$ is the actual domain size of the perturbed system, whereas $L(t)$ is the average domain size 
without a perturbation. As in the long time limit $L_p(t,s) \longrightarrow L(t)$, this quantity vanishes for
long observation times. The absolute values are used in Eq. (\ref{eq:M}) as we can have either
that
$L_p(s,s) > L(s)$ or that $L_p(s,s) < L(s)$, depending on whether $q_i > q_f$ or $q_i < q_f$.
In our study we considered multiple cases with various combinations of $q_i$ and $q_f$. In doing so,
we restricted ourselves to values of $q_i \ge 0.7$ as well as to 
not too large changes in $q$, such that $\left| q_i - q_f \right| \leq 0.1$.

Let us mention that the response $M(t,s)$ is a time integrated global response as (a) it sums up all the changes
that accumulate over the
time during which the perturbation is switched on and (b) it gives the global response of the system to a perturbation 
that affects all parts of the system in the same way. As such it is related in a rather complicated way to the 
response $R(t,s)$ discussed previously, which is the local response to an instantaneous perturbation.
It is not clear {\it a priori} whether a scaling form like that given in (\ref{eq:Rscaling})
remains valid for the more complicated response studied here.

%%%%%%%%%%%%%%%%%%%%%%%%%%%%%%%%%%%%%%%%%%%FIG 5.%%%%%%%%%%%%%%%%%%%%%%%%%%%%%%%%%%%%%%%%%%%%%%%%%%%%%%
\begin{figure} [h]
\includegraphics[width=0.80\columnwidth]{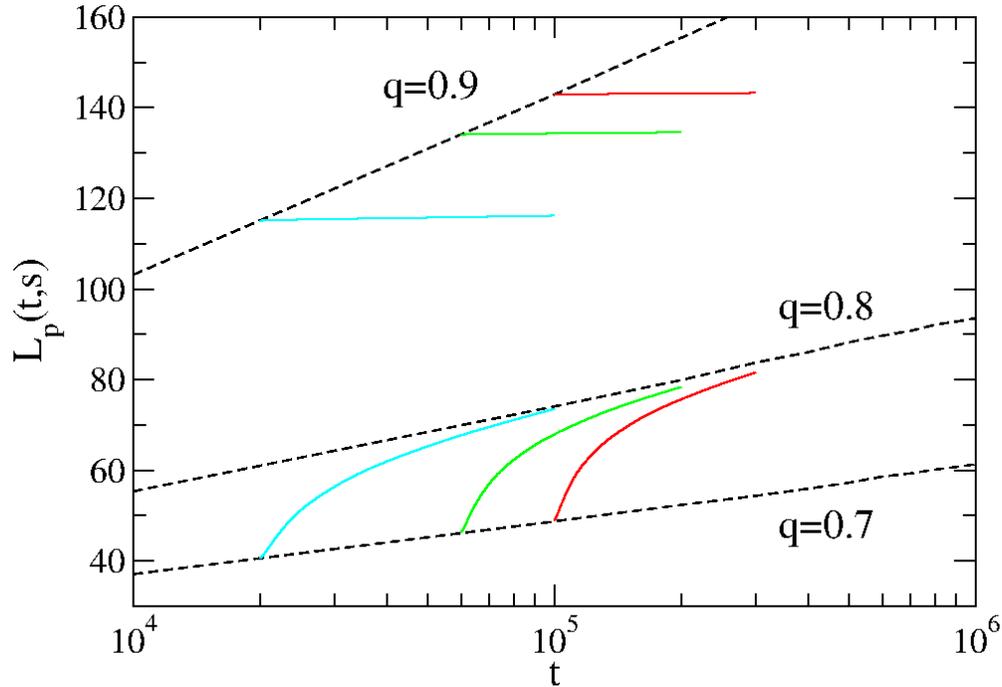}
\caption{\label{fig5} (Color online)
Time evolution of the average growth length when changing after the waiting time $s$ the value of the rate
$q$ from 0.9 to 0.8 (upper full colored lines) or from 0.7 to 0.8 (lower full colored lines). The different waiting times 
are s=20000 (cyan lines), s=60000 (green lines), and $s=100000$ (red lines).
}
\end{figure}
%%%%%%%%%%%%%%%%%%%%%%%%%%%%%%%%%%%%%%%%%%%FIG 5.%%%%%%%%%%%%%%%%%%%%%%%%%%%%%%%%%%%%%%%%%%%%%%%%%%%%%%

Let us start with a discussion of the time evolution of the domain length $L_p(t,s)$ after changing the value of the rate $q$.
As we see in Fig. \ref{fig5} for two cases with $q_f = 0.8$, the behavior of $L_p(t,s)$ is 
remarkably different depending on whether $q$ is decreased
or increased. When decreasing $q$ after the waiting time, see the upper colored curves in Fig. \ref{fig5},
the domain size is at the moment of the change much larger than the average domain size 
in the unperturbed system that evolves at the constant value $q = q_f$.
As a result domains grow extremely slowly after the change and it
takes a very long time for $L_p(t,s)$ to approach the unperturbed curve $L(t)$. A closer inspection reveals that
the difference $L_p(t,s) - L(s)$ varies logarithmically with time, $L_p(t,s) - L(s) =
\mu \ln t + \nu$, where $\mu$ is found to be independent of the waiting time $s$. 
The situation is very different for cases where
$q$ is increased, see the lower colored curves in Fig. \ref{fig5}. In these cases accelerated
growth sets in and the perturbed curve approaches the unperturbed curve very rapidly. Indeed, after an initial short time
regime, the difference between the two lengths $L_p(t,s)$ and $L(t)$ vanishes in an approximately algebraic way, with an effective exponent
whose value is between 1.7 and 1.9, depending on the waiting time $s$. 

%%%%%%%%%%%%%%%%%%%%%%%%%%%%%%%%%%%%%%%%%%%FIG 6.%%%%%%%%%%%%%%%%%%%%%%%%%%%%%%%%%%%%%%%%%%%%%%%%%%%%%%
\begin{figure} [h]
\includegraphics[width=0.80\columnwidth]{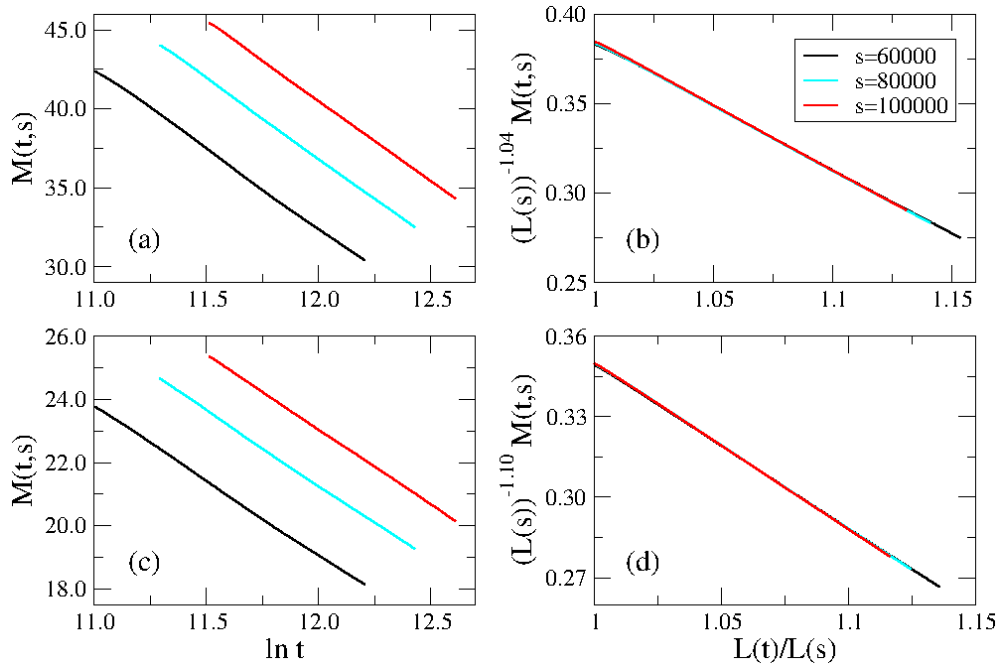}
\caption{\label{fig6} (Color online)
Response function for the $ABC$ model where at the waiting time $s$ the exchange rates are decreased
from some
initial value $q_i$ to the final value $q_f$: (a,b) $q_i=0.9$ and $q_f = 0.85$, (c,d) $q_i=0.8$ and $q_f = 0.7$.
Plotting the response function
against the scaling variable $L(t)/L(s)$, see (b) and (d), yields a perfect data collapse.
The data result from averaging over 10000 independent runs.
}
\end{figure}
%%%%%%%%%%%%%%%%%%%%%%%%%%%%%%%%%%%%%%%%%%%FIG 6.%%%%%%%%%%%%%%%%%%%%%%%%%%%%%%%%%%%%%%%%%%%%%%%%%%%%%%

\begin{figure} [h]
\includegraphics[width=0.80\columnwidth]{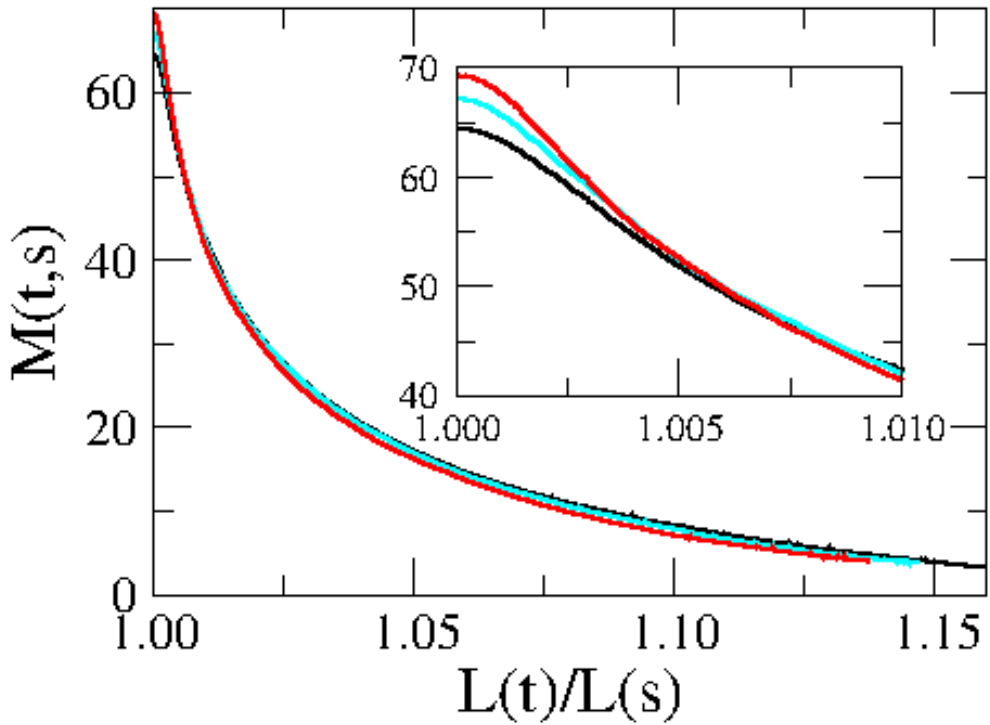}
\caption{\label{fig7} (Color online)
Response function for the $ABC$ model where at the waiting time $s$ the exchange rate is increased from the
initial value $q_i=0.8$ to the final value $q_f=0.9$. The waiting times are the same as in Fig.~\ref{fig5}.
As the different curves intersect, see inset, no data collapse can be achieved by simply multiplying $M(t,s)$
with a waiting time dependent constant.
The data result from averaging over 10000 independent runs.
}
\end{figure}

We investigate the possible scaling behavior of the response $M(t,s)$, see Eq. (\ref{eq:M}), in Figures \ref{fig6} and \ref{fig7}.
The case $q_i > q_f$ is illustrated in Fig. \ref{fig6} by two examples: a change from $q_i=0.9$ to
$q_f = 0.85$ as well as a change from $q_i=0.8$ to $q_f = 0.7$. We first remark, see Fig.\ \ref{fig6}a and \ref{fig6}c,
that $M(t,s)$ indeed varies linearly with $\ln t$, independent of the waiting time $s$. This observation already
suggests that the time integrated response also exhibits a scaling behavior where the time dependence is completely 
captured through the dynamic correlation length $L(t)$: 
\begin{equation}
M(t,s)  =  \left(L(s)\right)^{-\alpha} f_M\left(\frac{L(t)}{L(s)}\right)\label{eq:Mscaling}
\end{equation}
with the scaling variable $\frac{L(t)}{L(s)}$. As shown in Fig.\ \ref{fig6}b and \ref{fig6}d this indeed yields a
data collapse of the time integrated response, with an exponent $\alpha$ that depends on the rates $q_i$ and $q_f$:
$\alpha = 1.04(2)$ when changing the rate from 0.9 to 0.85 and $\alpha = 1.10(2)$ when changing the rate from 0.8 to 0.7.
It therefore follows that for the case $q_i > q_f$ the response shows a standard aging scaling, similar to the
autocorrelation, provided that the time-dependent length $L(t)$ is used.

%%%%%%%%%%%%%%%%%%%%%%%%%%%%%%%%%%%%%%%%%%%FIG 7.%%%%%%%%%%%%%%%%%%%%%%%%%%%%%%%%%%%%%%%%%%%%%%%%%%%%%%
%%%%%%%%%%%%%%%%%%%%%%%%%%%%%%%%%%%%%%%%%%%FIG 7.%%%%%%%%%%%%%%%%%%%%%%%%%%%%%%%%%%%%%%%%%%%%%%%%%%%%%%

This is completely different for the case $q_i < q_f$, see Fig. \ref{fig7}. As already discussed, the domains
at the moment of the change of the rate are smaller than those encountered in the unperturbed system
with the same number of time steps, and the larger rate $q_f$ yields a much higher probability
for a particle to jump from one domain to another. Consequently, the domain growth proceeds very fast. As shown in
Fig.\ \ref{fig7} for the case with $q_i=0.8$ and $q_f=0.9$, no good data collapse is observed 
when using as scaling variable $L(t)/L(s)$. In fact, see the inset,
the curves for different waiting times always cross, which of course renders a data collapse impossible. Clearly, when
the approach of $L_p(t,s)$ to $L(t)$ is faster than logarithmic, then a scaling behavior like that observed for
$q_i > q_f$ can not be expected. As mentioned before, $L(t) - L_p(t,s)$ displays in a certain regime an effective algebraic
dependence on $t$. This might suggest that we could choose as scaling variable $t/s$. However, as this effective exponent
displays a dependence on the waiting time, this also does not yield a data collapse.

Let us close this Section by mentioning a possible alternative way to probe the response of our system.
Adapting a protocol discussed in \cite{Hen12}, one can consider a space dependent rate where $q_x =
q_0 \pm a_x \, \varepsilon$ is the rate at position $x$. Here, $a_x = \pm 1$, whereas $\varepsilon$ is 
a small parameter. One would then consider two different realizations with the same noise (i.e. sequence 
of random numbers), one where the rate is kept fixed at $q = q_0$ and one where the space dependent rate 
$q_x$ is used up to the waiting time, after which the constant rate $q_0$ is used. Comparison of the resulting
configurations should then allow to monitor how the perturbed system relaxes toward the unperturbed system.
This alternative protocol is very close to the standard protocol used to calculate the
autoresponse in magnetic systems where a space
dependent random magnetic field is applied \cite{Bar98}. It remains to be seen, however, whether this approach 
allows one to sample the local response with good enough statistics. We leave it to a future study to 
clarify this point.

\section{Discussion and conclusion}

In recent years numerous studies have yielded a rather good understanding of aging processes governed by an algebraic growth of
the unique relevant length scale. This is especially true for systems with competing ground states where phase coarsening
dominates the out of equilibrium behavior in the ordered phase, thereby yielding a typical domain size that increases as a power-law
of time. Perfect magnets, as embodied by the Ising or Potts models, are well studied examples. However, as soon as one adds disorder
and/or frustration effects, the dynamics slows down. A series of recent numerical studies \cite{Par10,Cor11,Cor12,Par12} 
have confirmed the existence
of a crossover from an initial power-law like regime to an asymptotic regime where the relevant length scale increases
much slower with time. Even though it is expected that this long time regime is characterized by logarithmic growth,
none of the studies in which the time evolution of the system was followed 
were able to fully enter this asymptotic regime. Consequently,
most of the non-equilibrium relaxation properties in such a regime have not yet been explored.

Motivated by the absence of systematic studies of aging in system with logarithmic growth, we propose to follow a different route
and to focus on model systems for which it is possible to access the logarithmic regime. Even though these models are not related
to disordered magnetic systems, their study should allow us to gain a better understanding of the more universal properties encountered
in this regime.

In this paper we have studied the $ABC$ model and a related domain model with a simplified dynamics. The $ABC$ model allows us
to study the crossover from an early time regime to the logarithmic regime. The domain model, on the other hand, very rapidly
displays a logarithmic growth of the domains. Therefore, using this model we can test the scaling behavior of two-times quantities like
correlation and response functions. 

Our study shows that in the crossover regime the correlation function can be rather complicated. Once the domains are formed and
coarsening proceeds, one enters the logarithmic regime where for waiting times large enough the two-time autocorrelation starts
to exhibit a scaling behavior. This scaling behavior is fully elucidated when studying the domain model. In that case we find
for the autocorrelation function a standard aging scaling, provided that the time dependence is expressed through the length scale
$L(t)$ that increases logarithmically with time. 

In order to study the response of the system to a perturbation, we keep the swapping rate $q$, the only parameter
in the model, at some initial value $q_i$ up to the waiting time $s$, where we then change this rate and set it equal to
the final value $q_f$. We then compare the time evolution of the domains formed using this protocol with that of the domains that
are formed when from the start the rate is set equal to $q_f$. The response function is then a time integrated global response
to a global change in the system. Interestingly, we find different types of behavior, depending on whether the rate is decreased
or increased at the waiting time. If the rate is decreased, then the difference between the domain sizes of the perturbed and unperturbed
systems decreases logarithmically with time. This then yields again a simple aging scaling with the typical length $L(t)$ as scaling
variable, in complete analogy to the behavior of the autocorrelation function. This is completely different when considering the case
where $q$ is increased. In that case the domains of the perturbed system grow very fast and rapidly approach the size of the
unperturbed system, yielding a regime where the approach to the unperturbed regime displays an effective power-law behavior, with
effective exponents that depend on the waiting time. Consequently, no dynamical scaling is observed in that case.

We view the present study as a first step in the systematic study of aging properties of systems undergoing logarithmic growth.
We expect additional important insights through the study of space-time quantities, like the two-times space-time correlation
function. Also, up to now we restricted ourselves to the global response to a global change. In future, this should be 
extended to the investigation of the local response to a local perturbation.

The two models studied here have of course no direct relation with the magnetic systems that motivated our study. Still, we expect that
some of the results obtained in our study should also remain valid for magnetic systems with logarithmic growth. This is especially true
for the simple aging scaling with the scaling variable $L(t)/L(s)$ that is found for the autocorrelation.
We expect that this is a general feature of systems undergoing anomalous
slow dynamics that is characterized by a logarithmic growth of the typical domain size, including the
disordered ferromagnets. Future studies of other systems displaying
this type of growth should be able to substantiate this statement. Less obvious for us is whether the intriguing behavior 
encountered for the global response function is also a generic property. For the disordered ferromagnet the corresponding protocol
would consist in letting the system relax in the presence of a magnetic field $H$, whose value is then changed after the waiting
time (this final value could of course be $H=0$). We then should again have that the domains at the waiting time have 
a different typical length when compared with the domain size at constant magnetic field. The situation therefore seems
rather similar to what is discussed in this paper. Still, the domains in two- and three-dimensional ferromagnets are very different
to the pure domains encountered in the domain model. It therefore remains an intriguing question for the future whether
responses in other systems with anomalous slow dynamics behave in a similar way to what has been found in our study. 

\begin{acknowledgments}

This work is supported by the US National
Science Foundation through grants DMR-0904999 and DMR-1205309.

\end{acknowledgments}

\end{document}